\documentclass[
copyright,creativecommons]{eptcs}
\providecommand{\event}{CCA 2010} 
\usepackage{breakurl}             

\usepackage{amssymb}
\usepackage{latexsym}
\usepackage{amsfonts}
\usepackage{amsmath}
\usepackage{amssymb}

\newtheorem{thm}{Theorem}[section]

\newtheorem{prop}[thm]{Proposition}

\newtheorem{lemma}[thm]{Lemma}

\newtheorem{cor}[thm]{Corollary}

\newtheorem{Question}{Question}[section]
\newtheorem{definitiontemp}[thm]{Definition}
\newenvironment{defn}{\begin{definitiontemp}
\normalfont}{\end{definitiontemp}}

\newenvironment{pf}{\begin{trivlist}\item[\hskip\labelsep
{\it Proof.}]}{\end{trivlist}}

\newcommand{\qed}{\hbox to 0pt{}\nobreak\hfill\rule{2mm}{2mm}}

\newcommand{\Q}{\mathbb{Q}}
\renewcommand{\H}{\mathbb{H}}
\newcommand{\R}{\mathbb{R}}
\newcommand{\Z}{\mathbb{Z}}

\newcommand{\A}{\mathbb{A}}

\renewcommand{\P}{\mathcal{P}}

\newcommand{\Cbar}{\overline{C}}
\newcommand{\Sbar}{\overline{S}}

\def\phi{\varphi}

\newcommand{\la}{\langle}
\newcommand{\ra}{\rangle}

\newcommand{\ep}{\epsilon}

\newcommand{\pvec}{\vec{p}}
\newcommand{\xvec}{\vec{x}}

\newcommand{\yvec}{\vec{y}}
\newcommand{\Yvec}{\vec{Y}}
\newcommand{\zvec}{\vec{z}}
\newcommand{\set}[2]{\ensuremath{ \{ #1 : #2 \} }}

\title{The Cardinality of an Oracle in Blum-Shub-Smale Computation}
\author{Wesley Calvert
\institute{Murray State University\\
Murray, Kentucky 42071  USA}
\email{wesley.calvert@murraystate.edu}
\and
Ken Kramer \qquad\qquad Russell Miller
\institute{Queens College of CUNY\\
65-30 Kissena Blvd., Flushing, NY 11367 USA\\
CUNY Graduate Center\\
365 Fifth Avenue, New York, NY 10016 USA}
\email{kkramer@qc.cuny.edu \quad\qquad Russell.Miller@qc.cuny.edu}
}
\def\titlerunning{The Cardinality of an Oracle in BSS-Computation}
\def\authorrunning{Wesley Calvert, Ken Kramer, \& Russell Miller}
\begin{document}
\maketitle

\begin{abstract}
We examine the relation of BSS-reducibility on
subsets of $\R$.  The question was asked
recently (and anonymously) whether it is possible
for the halting problem $\H$ in BSS-computation
to be BSS-reducible to a countable set.
Intuitively, it seems that a countable set
ought not to contain enough information to decide
membership in a reasonably complex
(uncountable) set such as $\H$.  We confirm
this intuition, and prove a more general
theorem linking the cardinality of the oracle set
to the cardinality, in a local sense, of the set
which it computes.  We also mention other recent
results on BSS-computation and algebraic real numbers.
\end{abstract}

\section{Introduction}
\label{sec:intro}

Blum, Shub, and Smale introduced in \cite{BSS} a notion of computation
with full-precision real arithmetic, in which the ordered field
operations are axiomatically computable, and the computable functions
are closed under the usual operations.  A complete account of
this model is given in \cite{BCSS}.  A program for such a machine consists
of a finite set of instructions as described there, and the instructions
are allowed to contain finitely many real parameters, since a single real
number is viewed as a finite object.  The program can
add, multiply, subtract, or divide real numbers in its cells,
can copy or delete the content of a cell, and can
use the relations $=$ and $<$ to compare the contents of two cells,
forking according to whether the contents of those cells
satisfy that relation.  For our purposes, it will
be convenient to assume that the forking instructions in the program
compare the real number in a single given cell to $0$, under either
$=$ or $<$ or $>$.  Such a machine has equivalent computing power
to machines which can compare the contents of two different cells
to each other.

Of course, the BSS model is not the only concept of computation on $\R$,
nor should it be considered the dominant model.  It corresponds
to a view of the real numbers as a fixed structure, perhaps given
axiomatically --  defined, for instance, as the unique complete
ordered field, with field operations vouchsafed unto us mathematicians;
as opposed to a view of real numbers as objects defined by
Cauchy sequences or by Dedekind cuts in the rational numbers $\Q$,
with operations derived from the analogous operations on $\Q$.
There is no obvious method of implementing
BSS machines by means of digital computers.
This failure invites a contrast with computable analysis,
which treats real numbers as quantities
approximated by rational numbers and is intended to reflect the capabilities of
digital computers.   However, the BSS model is of interest
both for the analogy between it and the Turing model, which can be seen as
BSS computation on the ring $\Z/(2\Z)$, and because it reflects
the intuitions of many mathematicians -- dating back to the nineteenth
century, and mostly outside of computer science -- about
the notion of algorithmic computation on $\R$

This paper will consider sets of algebraic real numbers, and
other sets of tuples from $\R$, as oracles for BSS machines,
and will examine
the relative difficulty of deciding membership in such sets under the
BSS model of computation.  We will focus in particular on questions
about cardinality:  to what extent the complexity of a subset of $\R$
allows us to draw conclusions about its cardinality.  The previous paper \cite{MZ08}
by Meer and Ziegler focused attention on these issues, and here
we answer several of the questions raised there.  Our method
adapts a known technique from BSS computability, and should
be comprehensible to casual readers as well as to logicians
and computer scientists.  It requires significant use of algebraic
properties of the real numbers, in addition to computability,
reinforcing the general perception of the BSS model as an
essentially algebraic approach to computation on $\R$,
treating real numbers as indivisible finite items.
In contrast, the use of computable analysis normally results
in a more analytic approach to computation on $\R$.
As computable model theorists with experience in algorithms
on (countable) Turing-computable fields, we the present authors
are more familiar with the algebraic side.

Our notation generally follows that of \cite{MZ08}.
The set of all finite tuples of real numbers is denoted
$\R^\infty$; the inputs and outputs of BSS machines on $\R$
all lie in this set, and the collective content of the cells of a BSS
machine at a given stage in a computation may also be regarded
as an element of $\R^\infty$.  We use $\A$ to denote the set of all
real numbers which are algebraic over the subfield $\Q$
of rational numbers.  $\A$ is partitioned into subsets
$\A_{=d}$, for each $d\in\omega$:  $\A_{=d}$ contains those
algebraic real numbers of degree exactly $d$ over $\Q$.
(Recall that the \emph{degree} of $x$ over $\Q$ is the
vector space dimension over $\Q$ of the field $\Q(x)$
generated by $x$; equivalently, it is the degree of the
minimal polynomial of $x$ in $\Q[X]$.)  We also write
$\A_d=\cup_{c\leq d}\A_{=c}$, the set of algebraic real
numbers of degree $\leq d$.  By the definition of degree,
$\A_0$ is empty, and $\A_1$ contains exactly the rational
numbers themselves.  We mention \cite{vdW70} as an excellent
source for these and other algebraic preliminaries,
and \cite{FJ86} for more advanced questions about algorithms
on fields.

The following lemma is well known, and clear by induction on stages.
It reflects the fact that the four field operations are the only
operations which a BSS machine is able to perform.
\begin{lemma}
\label{lemma:fieldboundary}
If $M$ is a BSS machine using only the real parameters $\zvec$
in its program, then at every stage of the run of $M$ on any input
$\xvec$, the content of every cell lies in the field $\Q(\zvec,\xvec)$.
\qed\end{lemma}

It is immediate from this lemma that the set $\A$ of algebraic real numbers
cannot be the image of $\omega$ under any BSS-computable function,
as it is not contained within any finitely generated field. (Here $\omega$
represents the set of nonnegative integers, viewed as a subset of $\R$.)
We say that $\A$ is not \emph{BSS-denumerable}.  On the other hand,
$\A$ does satisfy the definition of \emph{BSS semidecidability},
which is the best analogue of Turing-computable
enumerability and has been studied more closely in the literature.
\begin{defn}
\label{defn:BSSsemidecidable}
A set $S\subseteq\R^\infty$ is \emph{BSS-semidecidable} if there
exists a (partial) BSS-computable function with domain $S$,
and \emph{BSS-denumerable} if there exists a partial BSS-computable
function mapping $\omega$ onto $S$.  $S$ is \emph{BSS-decidable}
if its characteristic function $\chi_S$ is BSS-computable.
\end{defn}
It is immediate that $S$ is BSS-decidable if and only if both
$S$ and $(\R^\infty -S)$ are BSS-semidecidable.  This justifies
the analogy between BSS-semidecidability in $\R^\infty$ and
computable enumerability in $\omega$, and also dictates
the use of the prefix ``semi.''  The term \emph{BSS-denumerable},
on the other hand, suggests that the set can be listed out,
element by element, by a BSS machine, which is precisely the
content of the definition above.  (The adjective \emph{denumerable}
was once a synonym for \emph{countable}, but has fallen out
of use in recent years.)  In the context of Turing computability,
computable enumerability and semidecidability are equivalent, but
in the BSS context, the set $\A$ distinguishes
the two notions, being BSS-semidecidable but not BSS-denumerable.
(On the other hand, every BSS-denumerable set
is readily seen to be BSS-semidecidable.)
The semidecision procedure for $\A$ is well-known:
take any input $x$, and go through all nonzero polynomials
$p(X)\in\Q[X]$, computing $p(x)$ for each.  If ever $p(x)=0$,
the machine halts.  The ability to go through the polynomials
in $\Q[X]$ follows from the BSS-denumerability of $\Q[X]$,
which in turn follows from the BSS-denumerability of $\Q$.
(A similar result applies to the set of algebraically dependent
tuples in $\R^\infty$; see for instance \cite{KZ08}.)

The question which gave rise to this paper was posed by Meer and Ziegler in
\cite{MZ08}.  (There they credit it to an anonymous referee of that paper.)
It uses the notion of a \emph{BSS reduction}, analogous to Turing reductions.
A \emph{oracle BSS machine} is essentially a BSS machine with the additional
ability to take any finite tuple (which it has already assembled on the cells of its tape),
ask an oracle set $A$ whether that tuple lies in $A$, and fork
according to whether the answer is positive or negative.
The oracle $A$ should be a subset of $\R^\infty$, of course,
and we will write $M^A$ to represent an oracle BSS program
(or machine) equipped with an oracle set $A$.
Oracle BSS programs can be enumerated (by tuples from
$\R^\infty$) in much the same manner as regular BSS programs.
If $B\subseteq\R^\infty$ and the characteristic function $\chi_B$
can be computed by an oracle BSS machine $M^A$ with oracle $A$,
then we write $A\leq_{BSS} B$, and say that $A$ is \emph{BSS-reducible}
to $B$, calling $M$ the \emph{BSS reduction} of $A$ to $B$.
Should $A\leq_{BSS} B$ and also $B\leq_{BSS}A$,
we write $A\equiv_{BSS}B$ and call the two sets
\emph{BSS-equivalent}.  All this is exactly analogous to oracle
Turing computation on subsets of $\omega$.

\begin{Question}\label{mzq} Let $\mathbb{A}$ be the set of algebraic numbers
in $\R$, i.e.\ those which are roots of a nonzero polynomial in $\Q[X]$.
Also, let $\H$ be the Halting Problem for BSS computation on $\R$,
as described in \cite[\S3.5]{BCSS}.  Is it true that $\H \not\leq_{BSS} \A$?
And more generally, could any countable subset of $\R^\infty$
contain enough information to decide $\H$?
\end{Question}

That $\A\leq_{BSS}\H$ is immediate.
Let $P$ be the BSS program which, on input $x\in\R$,
plugs $x$ successively into each nonzero polynomial $p(X)$
in (the BSS-denumerable set) $\Q[X]$ and halts if ever $p(x)=0$.
Then $x\in\A$ iff the program $P$ halts on input $x$.
(Similarly, every BSS-semidecidable set is BSS-decidable in $\H$,
and indeed $1$-reducible to $\H$ in the BSS model.)  The focus
of the question is on the lack of any reduction in the opposite direction.
Section \ref{sec:basic} gives the basic technical lemma
used in this paper to address such questions, and Section
\ref{sec:cardinality} applies it to give a positive answer to
Question \ref{mzq}.  We also prove
there a more general theorem relating BSS degrees to cardinality,
showing that for infinite subsets $S\subseteq\R$ and
$C\subseteq\R^\infty$, if $S\leq_{BSS} C$, then the local cardinality
(in a technical sense defined in that section) of $S$
cannot be greater than the (global, i.e.\ usual) cardinality of $C$.

\section{BSS-Computable Functions At Transcendentals}
\label{sec:basic}

Here we introduce our basic method for showing
that various functions on the real numbers fail to be
BSS-computable.  In Section \ref{sec:cardinality},
this method will be extended to give answers
about BSS-computability below certain oracles.
However, even the non-relativized version yields
straightforward proofs of several well-known
results about BSS-decidable sets, as we will
see shortly after describing the method.

In many respects, our method is equivalent to
the method, used by many others, of considering
BSS computations as paths through a finite-branching
tree of height $\omega$, branching
whenever there is a forking instruction in the program.
However, we think that the intuition for our method
can be more readily explained to a mathematician
unfamiliar with computability theory.
Our straightforward main lemma
says that near any transcendental input in its domain,
a BSS-machine must be defined by rational functions.
Where previous proofs usually made arguments about
countable sets of terminal nodes in the tree
of possible computations, we simply use
the transcendence of this element.
\begin{lemma}
\label{lemma:epnbhd}
Let $M$ be a BSS-machine, and $\zvec$ the finite tuple
of real parameters mentioned in the program for $M$.
Suppose that $\yvec\in\R^{m+1}$ is a tuple of real numbers
algebraically independent over the field $Q=\Q(\zvec)$,
such that $M$ converges on input $\yvec$.  Then there
exists $\ep>0$ and rational functions $f_0,\ldots,f_n\in Q(\Yvec)$,
(that is, rational functions of the variables $\Yvec$
with coefficients from $Q$) such that for all
$\xvec\in \R^{m+1}$ with $|\xvec-\yvec|<\ep$,
$M$ also converges on input $\xvec$ with output
$\la f_0(\xvec),\ldots,f_n(\xvec)\ra\in\R^{n+1}$.
\end{lemma}
\begin{pf}
The intuition is that by choosing $\xvec$ sufficiently close to $\yvec$,
we can ensure that the computation on $\xvec$ branches in
exactly the same way as the computation on $\yvec$,
at each of the (finitely many) branch points in the
computation on $\yvec$.  More formally,
say that the run of $M$ on input $\yvec$ halts at stage $t$,
and that at each stage $s\leq t$, the non-blank
cells contain the reals $\la f_{0,s}(\yvec),\ldots, f_{n_s,s}(\yvec)\ra$.
Lemma \ref{lemma:fieldboundary} shows that all
$f_{i,s}(\yvec)$ lie in the field $Q(\yvec)$, so each $f_{i,s}$
may be viewed as a rational function of $\yvec$ with
coefficients in $Q$.  Indeed, each rational function
$f_{i,s}$ is uniquely determined in $Q(\Yvec)$,
since $\yvec$ was chosen algebraically independent
over $Q$.

Let $F$ be the finite set
$$ F = \set{f_{i,s}(\Yvec)}{s\leq t~\&~i\leq n_s~\&~f_{i,s}\notin Q},$$
the set of nonconstant rational functions used in the computation.
Now for each $f_{i,s}\in F$, the preimage
$f_{i,s}^{-1}(0)$ is closed in $\R^{m+1}$,
and therefore so is the finite union
$$ U=\bigcup_{f_{i,s}\in F}~f_{i,s}^{-1}(0).$$
By algebraic independence, $\yvec$ does not lie in $U$,
so there exists an $\ep >0$ such that the $\ep$-ball
$B_\ep(\yvec)=\set{\xvec\in\R^{m+1}}{|\xvec-\yvec|<\ep}$,
does not intersect the closed set $U$,
and is contained within the domain of all $f_{i,s}\in F$.
This will be the $\ep$ demanded by the lemma.
Notice that more is true:  for all $f_{i,s}\in F$ and all
$\xvec\in B_\ep(\yvec)$, $f_{i,s}(\xvec)$ and
$f_{i,s}(\yvec)$ must have the same sign,
since otherwise there would be a path from
$\xvec$ to $\yvec$ within $B_\ep(\yvec)$, along
which $f_{i,s}$ would have to assume the value $0$.

Now fix any $\xvec\in B_\ep (\yvec)$.  We claim
that in the run of $M$ on input $\xvec$, at each stage $s\leq t$,
the cells will contain precisely $\la f_{0,s}(\xvec),\ldots,f_{n_s,s}(\xvec)\ra$
and the machine will be in the same state in which it was
at stage $s$ on input $\yvec$.  This is clear for
stage $0$, and we continue by induction, going
from each stage $s<t$ to stage $s+1$.  If the machine
executed a copy instruction or a field operation in this
step, then the result is clear, by inductive hypothesis.
Otherwise, the machine executed a fork instruction,
comparing some $f_{i,s}(\xvec)$ with $0$.  But we saw above
that $f_{i,s}(\xvec)$ and $f_{i,s}(\yvec)$ have the same sign
(or else $f_{i,s}(y)=0$, in which case $f_{i,s}$ is the constant function $0$),
so in both runs the machine entered the same state at
stage $s+1$, leaving the contents of all cells intact.
This completes the induction, and leaves us only to remark that
therefore, at stage $t$, the run of $M$ on input $\xvec$
must also have halted, with $\la f_{0,t}(\xvec),\ldots,f_{n,t}(\xvec)\ra$
in its cells as the output.
\qed\end{pf}

(If our BSS machines were allowed to compare the contents
of two cells under $=$ or $<$, as is standard, then
our set $F$ would have to consist of all nonconstant differences
$(f_{i,s}-f_{j,s})$.  The proof would still work, but the
method above is simpler.)

Lemma \ref{lemma:epnbhd} provides quick proofs of several
known results, including the undecidability of every
proper subfield $F\subset\R$.

\begin{cor}
\label{cor:decidable}
No BSS-decidable subset $S\subseteq \R^n$ can be both dense
and co-dense in $\R^n$.
\end{cor}
\begin{pf}
If the characteristic function $\chi_S$ were BSS-computable,
say by some machine $M$ with parameters $\zvec$,
then by Lemma \ref{lemma:epnbhd}, it would be constant
in some neighborhood of every $\yvec\in\R^n$ with coordinates
algebraically independent over $\zvec$.
\qed\end{pf}
Indeed, the same proof shows that any BSS-computable
total function with discrete image must be constant on each
of the $\ep$-balls given by Lemma \ref{lemma:epnbhd}.
\begin{cor}
\label{cor:bdry}
Define the boundary of a subset $S\subseteq\R^n$
to be the intersection of the closure of $S$ with the
closure of its complement.  If $S$ is BSS-decidable,
then there is a finite tuple $\zvec$ such that
every point on the boundary of $S$ has coordinates algebraically
dependent over $\zvec$.  In particular, if $M$ computes
$\chi_S$, then its parameters may serve as $\zvec$.
\end{cor}
\begin{pf}
This is immediate from Lemma \ref{lemma:epnbhd}.
\qed\end{pf}

Of course, Corollaries \ref{cor:decidable} and \ref{cor:bdry}
have been deduced long since from other known results,
in particular from the Path Decomposition Theorem described in \cite{BCSS}.
We include them here because of the simplicity of these proofs,
and because they introduce the methods to be used in the following section.

\section{Countable Oracle Sets}
\label{sec:cardinality}

It is natural to think of countability of a subset
$S\subseteq\R^\infty$ as a bound on the amount
of information which can be encoded into $S$.
This intuition requires significant restating before it can
be made into a coherent (let alone true) statement,
but we will give a reasonable version in this section.
In \cite{MZ08}, it was asked whether there could exist a countable
set $C\subseteq\R^\infty$ such that the halting problem $\H$
for BSS computation on $\R$ satisfies $\H\leq_{BSS} C$.
We will show that the answer to this question is negative.
For a formal definition of $\H$ in this context,
we refer the reader to \cite[\S3.5]{BCSS}.
Since it is equiconsistent with \textbf{ZFC}
for the Continuum Hypothesis to be false,
we will make our arguments applicable to all infinite cardinals
$\kappa <2^{\aleph_0}$, countable or otherwise.

First, of course, every subset of $\R^\infty$ is BSS-equivalent
to its complement, and so countability and co-countability
impose the same restriction on information content.
Of course, many sets of size continuum, with equally large complements,
are quite simple:  the set of positive real numbers,
for example, is BSS-decidable, hence less complex than
the countable set $\Q$ (cf.\ Corollary \ref{cor:decidable}).
So it is not possible to prove
absolute results relating cardinality and co-cardinality
(within $\R^\infty$) to BSS reducibility, but nevertheless,
we can produce theorems expressing the intuition
that countable sets are not highly complex in the BSS model.
This process will culminate in Theorem
\ref{thm:cardinality} below, but first we show that with a countable oracle,
one cannot decide the BSS halting problem $\H$.
We conjecture that $\H$ is not an upper bound on the degree of a countable set,
i.e.\ that such a set can still be BSS-incomparable with $\H$,
but no matter whether that conjecture holds or fails, it certainly constitutes progress
just to know that the upper cone of sets above $\H$ contains no countable sets.

\begin{thm}
\label{thm:haltingproblem}
If $C\subseteq\R^\infty$ is a set such that $\H\leq_{BSS} C$,
then $|C|=2^{\aleph_0}$.
\end{thm}
We note that by BSS-equivalence, these conditions also ensure
$|\R^\infty -C|=2^{\aleph_0}$,
and ensure $|\R^m-C|=2^{\aleph_0}$ whenever $C\subseteq\R^m$.

\begin{pf}
Let $C\subseteq\R^\infty$ have cardinality $<2^{\aleph_0}$,
and suppose that $M$ is an oracle BSS machine such that
$M^C$ computes the characteristic function of $\H$.  We fix a program
code number $p$ for the program which accepts inputs
$\la x_1,x_2\ra\in\R^2$, searches through nonzero polynomials $q$
in $\Q[Y_1,Y_2]$, and halts iff it finds one with $q(x_1,x_2)=0$.
Since the program coded by $p$ uses no real parameters,
$p$ may be regarded as a natural number, but in our
argument it can equally well be a tuple $\pvec$
from $\R^\infty$, with one or several real numbers
coding program parameters.  Then the elements of $C$, the
finitely many parameters $\zvec$ of $M$, and the parameters,
if any, in the program coded by $\pvec$ together generate
a field $E\subseteq\R$ which also has cardinality
$<2^{\aleph_0}$, and so $\R$ is an extension of infinite
transcendence degree (indeed of degree $2^{\aleph_0}$) over this $E$.
(Since $C\subseteq\R^\infty$, we need to be precise:
$E$ is generated by the coordinates $p_1,\ldots,p_j$ and $z_1,\ldots,z_k$
of the tuples $\pvec$ and $\zvec$, and the coordinates of each tuple in $C$.)
 
Now fix a pair $\la y_1,y_2\ra$ of real numbers algebraically
independent over $E$.  Hence $\la \pvec,y_1,y_2\ra\notin\H$,
so $M^C$ on this input halts after finitely many steps
and outputs $0$.  As in Lemma \ref{lemma:epnbhd}, we fix the finitely many
functions $f_{i,s}(\Yvec)\in E(Y_1,Y_2)$ such that $f_{i,s}(y_1,y_2)$
appears in the $i$-th cell at stage $s$ during this computation.
(The program code $\pvec\in E^\infty$ will stay fixed throughout
this proof, so we may treat it as part of the function $f_{i,s}$,
rather than as a variable.)
Let $F$ be the set of those functions $f_{i,s}$ which
are not constants in $E$, and fix an $\ep>0$
such that whenever $\la x_1,x_2\ra\in\R^2$ with $x_1\in B_\ep(y_1)$
and $x_2\in B_\ep(y_2)$, every $f\in F$ satisfies $f(x_1,x_2)\cdot f(y_1,y_2)>0$.
Write each $f\in F$ as a quotient $f=\frac{g}{h}$ with
$g,h\in E[Y_1,Y_2]$ in lowest terms, and let $n$ be the greatest
degree of $Y_2$ in all of these finitely many polynomials $g$ and $h$.

So far this mirrors the proof of Lemma \ref{lemma:epnbhd}, but an additional
condition is needed.  The oracle BSS machine $M^C$, running on input $\la \pvec,x_1,x_2\ra$,
can ask its oracle, at any stage $s$ and for any cell $i$, whether $f_{i,s}(x_1,x_2)$
lies in the oracle set $C$, and can fork according to the oracle answer.
So, in addition to choosing $\la x_1,x_2\ra$ within $\ep$ of $\la y_1,y_2\ra$,
we must ensure, for every $i$ and $s$, that
$[f_{i,s}(x_1,x_2)\in C\iff f_{i,s}(y_1,y_2)\in C]$.
On input $\la\pvec,y_1,y_2\ra$, we know by algebraic independence over $E$
that $f_{i,s}(y_1,y_2)\notin C$ unless $f_{i,s}$ is a constant function
(in which case $f_{i,s}(x_1,x_2)=f_{i,s}(y_1,y_2)$, of course).  So, for all of
the finitely many $f\in F$, we need to ensure that $f_{i,s}(x_1,x_2)\notin C$ as well.

Now choose $x_1\in\R$ to be transcendental over $E$
and within $\ep$ of $y_1$, and pick $x_2$
within $\ep$ of $y_2$ such that $x_2$ is 
algebraic over $\Q(x_1)$ but has degree $>n$ over $E(x_1)$.
For instance, let $x_1=y_1$ and $x_2=\sqrt[m]{x_1}+b$,
where $m>n$ is prime and $b\in\Q$ is selected to place
$x_2\in B_{\ep}(y_2)$.  It follows from \cite[Exercise 1, p.\ 256]{J85}
that the polynomial $(Y^m-x_1)$ is irreducible in the
one-variable polynomial ring $E(x_1)[Y]$, so this $x_2$
has degree $m$ over $E(x_1)$.

Thus, for any $f\in F$, if $a=f(x_1,x_2)\in E$,
then $0=g(x_1,x_2)-ah(x_1,x_2)$.  Since $f$ is nonconstant,
$g$ is not a scalar multiple of $h$, and so $(g-ah)$ would then
be a nonzero polynomial in $E[Y_1,Y_2]$ of degree $\leq n$,
contradicting our choice of $x_2$.  Hence $f(x_1,x_2)\notin E$
for every $f\in F$.  But then the oracle computation
$M^C(\pvec,x_1,x_2)$ must follow the same path as $M^C(\pvec,y_1,y_2)$
and give the same output, namely $0$.  Since $\la \pvec,x_1,x_2\ra\in\H$,
this proves that $M^C$ does not compute the characteristic function of $\H$.
\qed\end{pf}

Indeed the preceding proof shows more than was stated.
\begin{cor}
\label{cor:subfield}
If $C\subseteq\R^\infty$ is a set such that $\H\leq_{BSS} C$,
then $\R$ has finite transcendence degree over the field
$K$ generated by (the coordinates of the tuples in) $C$,
and also has finite transcendence degree over the field
generated by the complement of $C$.
\end{cor}
\begin{pf}
Given an oracle BSS machine $M$ which computes $\H$ from oracle $C$,
let $E$ be the extension field $K(\zvec,\pvec)$, with $K$ as defined in the
corollary.  If $\R$ had transcendence degree $\geq 2$ over this $E$,
then the proof of Theorem \ref{thm:haltingproblem}
would go through:  we could choose $y_1,y_2\in\R$
algebraically independent over $E$, say with $y_1>0$,
and again let $x_1=y_1$ and $x_2=b+\sqrt[m]{x_1}$,
with $m$ and $b$ as in that proof.
But this would show that $M^C$ does not compute $\H$.
So $\R$ has transcendence degree $\leq 1$ over
this $E$, and therefore is algebraic over $E(t)=K(t,\zvec,\pvec)$
for some $t\in\R$.

Since $C$ is BSS-equivalent to its complement,
the same proof applies to $(\R^\infty-C)$,
and also to $(\R^m-C)$ if $C\subseteq\R^m$.
\qed\end{pf}

As we consider the general case of a BSS computation
of the characteristic function $\chi_S$ of a set $S\subseteq\R$
using an oracle $C$ of infinite cardinality $\kappa<2^{\aleph_0}$,
the following definition will be useful.  Here $\Sbar$ denotes
$(\R-S)$, the complement of $S$ in $\R$
(as opposed to the topological closure).
\begin{defn}
\label{defn:bicardinality}
A set $S\subseteq\R$ is \emph{locally of
bicardinality $\leq\kappa$} if there exist
two open subsets $U$ and $V$ of $\R$
with 
$|\R-(U\cup V)|\leq\kappa$
and $|U\cap S|\leq \kappa$ and
$|V\cap \overline{S}|\leq \kappa$.

The \emph{local bicardinality of $S$} is the least cardinal
$\kappa$ such that $S$ is locally of bicardinality $\leq\kappa$.
\end{defn}

If $\kappa<2^{\aleph_0}$, then such $U$ and $V$ must be disjoint, since
$(U\cap V)$ is open with $|U\cap V|
\leq |U\cap S|+|V\cap\Sbar|\leq\kappa$.
So the definition roughly says that up to sets of size $\kappa$,
each of $S$ and $\Sbar$ is equal to an open subset of $\R$.
In Lemma \ref{lemma:Cantorbicard} below,
we will show that the Cantor middle-thirds set
has local bicardinality $2^{\aleph_0}$.

The property of local bicardinality $\leq\kappa$ does not appear
to us to be equivalent to any more easily stated property,
and we are not aware of it having been used (or even
stated) elsewhere in the literature.
The same definition in higher dimensions
completely loses its power: any connected component $U_0$
of $U$ must have boundary $\partial U_0$
with $U_0\cap\partial U=V\cap\partial U_0=\emptyset$,
since $U$ and $V$ are open and disjoint.  But then
$|\partial U_0|\leq |\R^n-(U\cup V)|\leq\kappa$,
which is feasible in $\R^1$ but not in higher dimensions,
unless $U$ or $V$ were empty or $\kappa=2^{\aleph_0}$.
Thus, in $\R^n$ with $n>1$, every set of local bicardinality
$<2^{\aleph_0}$ has either cardinality $<2^{\aleph_0}$
or co-cardinality $<2^{\aleph_0}$.
Nevertheless, within $\R^1$, this is exactly the condition
needed in our general theorem on cardinalities.

\begin{thm}
\label{thm:cardinality}
If $C\subseteq\R^\infty$ is an oracle set of
infinite cardinality $\kappa <2^{\aleph_0}$, and
$S\subseteq\R$ is a set with $S\leq_{BSS} C$, 
then $S$ must be locally of bicardinality $\leq\kappa$.
The same holds for oracles $C$ of infinite co-cardinality
$\kappa<2^{\aleph_0}$.
\end{thm}
\begin{pf}
Again let $\zvec$ be the parameters used by the oracle
BSS machine $M$ which, given oracle $C$, computes $\chi_S$.
Then for any input $y\in\R$ transcendental over the subfield
$E$ of cardinality $\kappa$ generated by $\zvec$ and the individual
coordinates of all elements of $C$, there will again exist a finite set
$F_y\subseteq E(X)$ as above, and an $\ep>0$ such that $f(x)\cdot f(y)>0$
for all $x\in B_\ep (y)$ and $f\in F_y$.  For each such $y$,
let $B(y)$ be an open interval of length less than
the corresponding $\ep$, such that $B(y)$ contains $y$
and has rational end points.
Now if $x\in B(y)$ is also transcendental over $E$, then the
computation of $\chi_S(x)$ using this machine and the
$C$-oracle proceeds along the same path as the
computation for $y$, since $f(x)\notin E$ for all $f\in F_y$.
(Indeed, this would hold whenever $x\in B(y)$
has degree $>n$ over $E$, where $n$ is the maximum
degree of all numerators and denominators of elements of $F_y$.)
This shows that $\chi_S(x)=\chi_S(y)$
for all such $x$.  Since only $\kappa$-many elements
of $B(y)$ can be algebraic over the
size-$\kappa$ field $E$, it follows that either
$|S\cap B(y)|\leq\kappa$ (if $y\notin S$) or
$|\overline{S}\cap B(y)|\leq\kappa$ (if $y\in S$).

Now if $t\in B(y_0)\cap B(y_1)$ with $t$, $y_0$, and
$y_1$ each transcendental over $E$,
then $t$ follows the same computation path as both $y_0$ and $y_1$,
implying that $\chi_S(y_0)=\chi_S(y_1)$ whenever $B(y_0)\cap B(y_1)\neq\emptyset$,
and therefore that either $B(y_0)\cap S$ and $B(y_1)\cap S$
both have size $\leq\kappa$, or else $B(y_0)\cap\Sbar$
and $B(y_1)\cap\Sbar$ both have size $\leq\kappa$.
So when we set
$$ U=\bigcup\set{B(y)}{|S\cap B(y)|\leq\kappa}
\text{~~~and~~~} 
V=\bigcup\set{B(y)}{|\Sbar\cap B(y)|\leq\kappa},
$$
we will have $U\cap V=\emptyset$.  Here the
unions are over those $y\in\R$ transcendental over $E$
(as $B(y)$ is not defined for any other $y$),
and so the complement $\R-(U\cup V)$ is a subset
of the algebraic closure of $E$, which has size $\kappa$.
Moreover, being a union of open intervals $B(y)$ with rational
end points, $U$ in fact equals the union of countably
many such intervals, say $U=\cup_{i\in\omega} B(y_i)$
for some sequence $y_0,y_1,\ldots$.  Since each
$B(y_i)$ has intersection of size $\leq\kappa$ with $S$
(and since $\kappa\geq{\aleph_0}$),
so does the entire union $U$.  Likewise
$|\Sbar\cap V|\leq\kappa$, proving the theorem.


The claim about oracles of co-cardinality $\kappa$
follows from applying the same argument to the
oracle $(\R^\infty-C)$, which is BSS-equivalent to $C$.
If $C\subseteq\R^m$ for some $m$, then the same
holds of $(\R^m-C)$.
\qed\end{pf}

Notice that the set $S$ of smaller complexity must be a subset of $\R$,
whereas $C$ is allowed to contain tuples from $\R^\infty$.
We conjecture that to extend the theorem to sets
$S\subseteq \R^\infty$, we would need to allow
$\R^\infty-(U\cup V)$ to be a union of $\kappa$-many
proper algebraic varieties defined over the field
generated by $C$.  
It is an open question (of interest only under
$\neg\textbf{CH}$) whether it is equivalent, for the purposes
of this conjecture and Theorem \ref{thm:cardinality},
to replace $|\R^\infty-(U\cup V)|\leq\kappa$
by $|\R^\infty-(U\cup V)|\leq{\aleph_0}$ here
or in Definition \ref{defn:bicardinality}.

To understand that this theorem cannot readily be stated
using a simpler property than Definition \ref{defn:bicardinality},
consider the BSS-computable set
$$ S=\set{x\in (0,1)}{(\exists m\in\omega)~2^{-(2m+1)}\leq x\leq 2^{-(2m)}},$$
containing those $x\in (0,1)$ which have a binary expansion
beginning with an even number of zeroes.
Then clearly no open interval $B$ which is locally of
bicardinality $\leq\kappa <2^{\aleph_0}$
can contain any of the countably many points $2^{-m}$,
so the theorem cannot require the complement
$\R^\infty-(U\cup V)$ to be finite, let alone empty.
Moreover, every open interval $B\subseteq\R$
which either contains $0$ or has left end point $0$
must have intersection of size $2^{\aleph_0}$ with both
$S$ and $\overline{S}$.  One can make the same
happen not only at $0$, but at each rational in
a sequence approaching $0$, and with such tricks one can
create examples defying most conceivable simplifications
of Theorem \ref{thm:cardinality}.

\section{The Cantor Set}
\label{sec:Cantor}

As an example of a set of local bicardinality $2^{\aleph_0}$,
we consider the Cantor set $C$, well known as
a set of measure $0$ within $\R$ which nevertheless
has cardinality $2^{\aleph_0}$.  By definition, $C$ contains
all real numbers $x\in [0,1]$ having ternary expansions in
only $0$'s and $2$'s.  One usually views $C$
as the set of numbers in the unit interval $[0,1]$
which remain after $\omega$-many
iterations of deleting the open ``middle third'' of each interval
(starting with the middle third $(\frac13,\frac23)$ of $[0,1]$).
It is clear from this description that $C$ is co-semidecidable
in the BSS model:  even a Turing machine can enumerate
all those middle-third intervals to be deleted.
Hence $\Cbar\leq_{BSS}\H$
(indeed via a $1$-reduction), forcing $C\leq_{BSS}\H$ as well.
The natural next question, whether $\H\leq_{BSS} C$,
was settled in \cite{Y08}, as described below.

\begin{lemma}
\label{lemma:Cantor}
The Cantor set $C$ is not BSS-semidecidable.
\end{lemma}
\begin{pf}
Since $\Cbar$ is semidecidable, semidecidability of $C$
would show that $C$ was BSS-decidable.  However, for every BSS-machine
with finite parameter tuple $\zvec$, $C$ contains some $y$
transcendental over $\Q(\zvec)$, since otherwise $C$ would be countable.
Now no nonempty open interval within $\R$ is contained within $C$,
and so every $\ep$-ball around $y$ contains elements of $\Cbar$.
Lemma \ref{lemma:epnbhd} therefore shows that $M$ does not
compute the characteristic function $\chi_C$.
\qed\end{pf}

The next lemma, combined with Theorem \ref{thm:cardinality},
would also immediately prove Lemma \ref{lemma:Cantor}.
On the other hand, it
dashes the hope that Theorem \ref{thm:cardinality} might prove
$\H\not\leq_{BSS} C$ the same way it proved $\H\not\leq_{BSS}\A$.
\begin{lemma}
\label{lemma:Cantorbicard}
The Cantor set $C$ has local bicardinality $2^{\aleph_0}$.
\end{lemma}
\begin{pf}
Suppose $C$ were locally of bicardinality $\leq\kappa <2^{\aleph_0}$.
Then we would have open disjoint sets $U$ and $V$
satisfying Definition \ref{defn:bicardinality}, and $C$, having size
$2^{\aleph_0}$, would have to intersect $V$ in some point $x$, since
$$ C-V\subseteq (U\cap C)\cup \overline{(U\cup V)}$$
and the right-hand side has size $\leq\kappa$.  The open set $V$
would then contain an $\ep$-ball around $x$.
However, every open interval around $x$
intersects each of $C$ and $\Cbar$ in $2^{\aleph_0}$-many points.
(To see this, just consider all $y$ whose ternary expansions match
that of $x$ for sufficiently many places to lie within that interval.)
Therefore $|V\cap\Cbar |=2^{\aleph_0}$, yielding a contradiction.
\qed\end{pf}
\begin{cor}
The Cantor set $C$ is not BSS-semidecidable below $\A$,
or below any other oracle of cardinality $<2^{\aleph_0}$.
\end{cor}
\begin{pf}
This simply means that no function which is BSS-computable in
the oracle $\A$ can have $C$ as its domain.  Indeed,
if it did, then $C\leq_{BSS}\A$, since $C$ and $\Cbar$ would both be
$\A$-semidecidable.  Lemma \ref{lemma:Cantorbicard} and Theorem
\ref{thm:cardinality} together rule out this possibility.
The same holds for any other oracle of size $<2^{\aleph_0}$.
\qed\end{pf}

Corollary \ref{cor:subfield}, our other natural hope for proving $\H\not\leq_{BSS}C$,
also fails to do so,
for the field generated by $C$ does not satisfy the hypothesis there.
It seems counterintuitive that a set of measure $0$ could generate
such a large field, so we prove it here.  (The authors assume that
this fact has been proven long since, and would appreciate a reference for it.)
\begin{lemma}[Folklore]
\label{lemma:generateR}
The Cantor set $C$ generates the entire field $\R$.
Indeed, it generates $\R$ as a ring.
\end{lemma}
\begin{pf}
The argument is best understood by seeing an example.  Here we
begin with an element of $[0,1]$, chosen arbitrarily, in ternary form:
\begin{align*}
&0.2201020001211\ldots\\
=~~&0.2200020000200\ldots\\
+ &0.0001000001011\ldots\\
=~~&0.2200020000200\ldots\\
+ (&0.0002000002022\ldots)\cdot\frac12
\end{align*}
Since $\frac12$ lies in every subfield of $\R$, this shows
that this number is generated from $C$ by field operations.
Indeed, since $\frac12=2\cdot\frac14=2\cdot (0.020202\ldots)$,
the number is generated from elements of $C$ by ring operations.
The same process can be applied to any element of $[0,1]$,
so $C$ generates the entire unit interval, and hence all of $\R$.
\qed\end{pf}

At this point the authors abandoned their search for a proof that
$\H\not\leq_{BSS}C$.  Fortunately, an anonymous referee
familiar with Yonezawa's paper \cite{Y08} pointed out the
necessary result there.
\begin{thm}[Corollary 2.5 in \cite{Y08}]
\label{thm:Yonezawa}
The sets $\Q$ and $C$ are BSS-incomparable.
\qed\end{thm}
Since the BSS-semidecidable set $\Q$ must be $\leq_{BSS}\H$,
this immediately answers the question.
\begin{cor}
\label{cor:Yonezawa}
$\H\not\leq_{BSS}C$.
\qed\end{cor}

\section{Other Results}
\label{sec:other}

In addition to the theorems on cardinality described above,
the authors have proven a selection of results on BSS-reducibility
among the different sets $\A_{=d}$, where
$$ \A_{=d} =\set{x\in\R}{x\text{~is algebraic over $\Q$ with minimal polynomial of degree~}d}.$$
For reasons of space, we omit most discussion of these theorems here,
as well as their proofs.  (They were presented by the third author in
a short talk at the meeting \emph{Logical Approaches to Computational
Barriers} in Greifswald, Germany in February 2010.)
However, we do state the main theorems here.
The basic result simply concerns $\A_{=d-1}$ and $\A_{=d}$,
and a proof appears in \cite{CKM10}.

\begin{thm}
\label{thm:general}
For every $d>0$, $\A_{=d}\not\leq_{BSS}\A_{d-1}$.
\end{thm}

This is generalized to a pair of arbitrary degrees.
Neither direction is trivial, but when $p$ is prime to $\frac{r}{p}$,
the backwards direction is implicit in \cite{MZ08}, by Meer and Ziegler,
and one particular case is explicitly shown by them.
\begin{thm}
\label{thm:allpq}
Let $p$ and $r$ be any nonnegative integers.
Then $\A_{=p}\leq_{BSS}\A_{=r}$ if and only if $p$ divides $r$.
\end{thm}
Of course, $\A_{=0}$ is just the empty set,
and $\emptyset\lneq_{BSS}\A_{=d}$ for all $d>0$,
since Meer and Ziegler showed in \cite{MZ08} that
no $\A_{=d}$ with $d>0$ is BSS-decidable.
So the theorem also holds when $p=0$,
but not when $p>0=r$.

To extend these results further, we define, for all $S\subseteq\omega$,
$\A_S = \cup_{d\in S}\A_{=d}$, the set of all algebraic
real numbers whose degrees over $\Q$ lie in $S$.
The proof of Theorem \ref{thm:general} is readily adjusted to yield
the following.
\begin{thm}
\label{thm:full}
For every $d>0$ in $\omega$ and every set $S\subset \omega$ with
$S\cap d\Z=\emptyset$, $\A_{=d}\not\leq_{BSS}\A_{S}$.
\end{thm}
\begin{cor}
\label{cor:lattice}
Let $P$ be the set of all prime numbers in $\omega$.
Then for all $S$ and $T$ in the power set $\P (P)$,
$\A_S \leq_{BSS} \A_T$ if and only if $S \subseteq T$.
\end{cor}
An immediate further corollary imparts substantial
richness to the partial order of the BSS-semidecidable degrees.
\begin{cor}
\label{cor:lattice2}
There is a subset $\mathcal{L}$ of the BSS-semidecidable
degrees such that $(\mathcal{L},\leq_{BSS})\cong (\mathcal{P}(\omega),\subseteq)$.
\end{cor}
\begin{pf}
We have $(\mathcal{P}(\omega),\subseteq)\cong(\mathcal{P}(P),\subseteq)$,
and Corollary \ref{cor:lattice} shows that the latter partial order
embeds into the BSS-semidecidable degrees via the
map $S\mapsto\A_S$.
\qed\end{pf}
We emphasize that Corollary \ref{cor:lattice2} only states that
there exists an isomorphism between the two partial orders.
It is unknown whether this map is also an isomorphism
of the two structures as lattices, or indeed whether
an arbitrary $\A_S$ and $\A_T$ must have a
greatest lower bound under $\leq_{BSS}$.
Of course, for $S,T\subseteq P$, $\A_{S\cap T}$ is the obvious candidate,
and if it really were the greatest lower bound, we would have
many \emph{minimal pairs} of BSS-semidecidable degrees.
(Recall that in Turing computability, a \emph{minimal pair}
consists of two degrees $\bf{c}$ and $\bf{d}$ whose infimum
is the computable degree $\bf{0}$.  The existence of a minimal
pair of nonzero computably enumerable degrees was
a significant result in Turing computability.)

Finally, we consider reducibility among the sets $\A_S$ and $\A_T$,
for arbitrary $S,T\subseteq\omega$.  Certain questions here
remain open.  First, we have a negative result.
\begin{thm}
\label{thm:SandT}
For sets $S,T\subseteq\omega$,
if $\A_S\leq_{BSS}\A_T$, then there exists
$N\in\omega$ such that all $p\in S$ satisfy
$\{ p,2p,3p,\ldots,Np\}\cap T\neq\emptyset$.
\end{thm}

The next two propositions are both positive results
(showing that reducibilities do exist).  Proposition \ref{prop:finiteS-T}
uses a nonuniform construction, and therefore only
applies when the set-theoretic difference $(S-T)$
is finite.  Proposition \ref{prop:relprime}
has a nonuniform construction, but requires a stronger hypothesis
involving relative primality.

\begin{prop}
\label{prop:finiteS-T}
For any subsets $S$ and $T$ of $\omega$,
if $(S-T)$ is finite and for every $p\in S-T$,
there exists an integer $q>0$ such that
$pq\in T$, then $\A_S\leq_{BSS}\A_T$.
\end{prop}

\begin{prop}
\label{prop:relprime}
Let $S$ and $T$ be subsets of the positive integers.
Suppose that for some absolute constant $N$ and each $d \in S$,
there is a positive integer $n_d \le N$ and prime to $d$ such that
$dn_d \in T$.  Then $\A_S \le_{\rm BSS} \A_T$.
\end{prop}
Of course, $n_d$ is allowed to equal $1$, since $1$ is prime
to $d$.  Thus every element of $S\cap T$ is immediately accounted
for, and only elements of $(S-T)$ can pose problems.
When $(S-T)$ is finite, Proposition \ref{prop:finiteS-T}
handles those problems, showing how to prove the result
even in the absence of relative primality.
When $(S-T)$ is infinite, the proof of Proposition \ref{prop:finiteS-T}
no longer applies.
One would hope to be able to remove from Proposition
\ref{prop:relprime} the assumption that $n_d$
must be prime to $d$, or else to extend 
Theorem \ref{thm:SandT} to yield a nonreducibility
result for this case, but for now this problem remains open.


\begin{thebibliography}{99}
\bibitem{BCSS}
L.\ Blum, F.\ Cucker, M.\ Shub, and S.\ Smale;
\emph{Complexity and real computation}
(Berlin:  Springer-Verlag, 1997).


\bibitem{BSS}
L.\ Blum, M.\ Shub, and S.\ Smale;
On a theory of computation and complexity over the real numbers,
\emph{Bulletin of the American Mathematical Society (New Series)}
\textbf{21} (1989), 1--46.

\bibitem{CKM10}
W.\ Calvert, K.\ Kramer, \& R.\ Miller;
Noncomputable functions in the Blum-Shub-Smale model,
in the abstract booklet for the conference \emph{Logical Approaches to Barriers in
Computing and Complexity} (17-20 February 2010, Greifswald, Germany).
Available at \texttt{qcpages.qc.cuny.edu/$^{\sim}$rmiller/BSSabstract.pdf}.


\bibitem{FJ86}
M.D.\ Fried \& M.\ Jarden;
\emph{Field Arithmetic}
(Berlin:  Springer-Verlag, 1986).

\bibitem{G08}
C.\ Gassner;
A hierarchy below the halting problem for additive machines,
\emph{Theory of Computing Systems} 
\textbf{43} (2008) 3--4, 464--470.


\bibitem{J85}
N.\ Jacobson;
\emph{Basic Algebra I}
(New York:  W.H.~Freeman \& Co., 1985).

\bibitem{KZ08}
W.\ Koolen \& M.\ Ziegler;
Kolmogorov complexity theory over the reals,
in \emph{Proceedings of the Fifth International Conference on
Computability and Complexity in Analysis, CCA '08},
\emph{Electronic Notes in Theoretical Computer Science}
\textbf{221} (Elsevier, 2008), 153-169.






\bibitem{MZ08}
K.\ Meer and M.\ Ziegler;
An explicit solution to Post's Problem over the reals,
\emph{Journal of Complexity} 
\textbf{24} (2008) 3--15.



\bibitem{vdW70}
B.L.\ van der Waerden;
\emph{Algebra}, volume I,
trans.\ F.\ Blum \& J.R.\ Schulenberger
(New York:  Springer-Verlag, 1970 hardcover, 2003 softcover).

\bibitem{Y08}
Y.\ Yonezawa;
The Turing degrees for some computation model
with the real parameter,
\emph{J.\ Math.\ Soc.\ Japan}
\textbf{60} 2 (2008), 311-324.


\end{thebibliography}
\end{document}